# Exciton Annihilation by Lanthanide Dopants: An Atomic Probe of Sub-Diffraction Exciton Diffusion in Ferromagnetic CrI$_3$

Kimo Pressler, Daniel R. Gamelin*

*Department of Chemistry, University of Washington, Seattle, WA 98195, United States*

Email: *gamelin@uw.edu*

**Abstract.** Excitons in two-dimensional (2D) magnetic van der Waals (vdW) materials offer unique windows into the properties of strongly correlated electrons. Their generation can be used to drive magnetic phase transitions, manipulate spins coherently, or access novel non-equilibrium regimes. Despite extensive investigation into the spin physics of CrI$_3$, exciton dynamics in this archetypal magnetic 2D material remain underexplored. Here, we report the use of Yb$^{3+}$ impurity point defects as exciton annihilators to probe exciton diffusion in CrI$_3$. Variable-temperature photoluminescence (PL) measurements for a series of $x$% Yb$^{3+}$:CrI$_3$ samples reveal thermally activated Dexter-type site-to-site hopping of excitons, with low exciton diffusivity associated with strong electron-nuclear coupling. Using Monte Carlo modeling calibrated by the experimental data, diffusivities are found to be orders of magnitudes lower than in 2D vdW semiconductors. Exciton diffusion lengths ($L_D$) are below ~3 nm at all temperatures, and thus well below the optical diffraction limit. These results have basic implications for the use of excitons to probe and manipulate their surroundings in this and related magnetic CrX$_3$ materials.

The recent demonstration of stable long-range magnetic order in monolayer CrI$_3$[1] has paved the way for new technologies based on ferromagnetic van der Waals (vdW) materials[2-5] and sparked renewed interest in the fundamental properties of CrI$_3$ and related compounds.[6-8] Despite its intriguing magnetic properties, one aspect of CrI$_3$ photophysics that has not yet been sufficiently investigated is how photogenerated excitons propagate within the lattice in space and time. Real-space migration of excitons is central to many existing and future optoelectronic technologies. Exciton diffusion in CrI$_3$ is essential to proposed schemes for using exciton scattering to control topological magnetic structures using light or for generating an excitonic anomalous Hall effect.[9,10] Likewise, exciton diffusion to surfaces is suggested to enable a surface-localized defect-assisted Auger recombination process that enhances lattice magnetization in CrCl$_3$.[11] A more complete description of exciton diffusion in CrX$_3$ materials may advance the basic-science understanding of such phenomena and aid efforts to harness their photophysics in future spin-based optoelectronic technologies.

Whereas spatial probes of exciton diffusion in vdW materials often employ time-resolved optical microscopy,[12-14] such methods are constrained by optical diffraction limits in both excitation and detection, with typical resolution of ~500 nm in the near-IR. Super-resolution techniques can improve upon this limitation by achieving bin sizes of 10s of nanometers or smaller, but this may still be insufficient for probing exciton dynamics in a material like $CrI_3$, because unlike many other workhorse vdW materials such as transition-metal dichalcogenides, $CrI_3$ excitons are localized around individual $Cr^{3+}$ ions. Indeed, the optical responses of $CrX_3$ (X = $Cl^-$, $Br^-$, $I^-$) compounds have historically been described using the ligand-field model, which thoroughly accounts for their multitude of absorption features as $d$-$d$ and ligand-to-metal charge-transfer (LMCT) transitions centered at individual $CrX_6^{3-}$ pseudo-octahedra.[15-17] Although recent reports have variously referred to the emissive excited states of $CrX_3$ compounds as exciton polarons,[18-20] Frenkel excitons,[21-23] or self-trapped excitons (STEs),[24] all of these interpretations invoke strong localization. Computational studies have also suggested that $CrI_3$ excitons localize at individual $Cr^{3+}$ ions.[21, 25] Direct optical probes are thus poorly suited for quantifying $CrI_3$ exciton diffusion.

An alternative approach to probing exciton diffusion is to monitor the annihilation of excitons under various conditions. For example, recent experiments have used variable-excitation-rate photoluminescence (PL) measurements in $CrX_3$ compounds to demonstrate exciton–exciton annihilation,[19, 26, 27] an Auger-like process in which two excitons combine energies to generate just one high-energy exciton. From excitation statistics and PL data, information about the exciton diffusion preceding annihilation can be deduced, and such experiments have been used to estimate a room-temperature exciton diffusion coefficient ($D$) between ~$10^{-6}$ and ~$10^{-2}$ cm$^2$/s for bulk $CrCl_3$.[27] This approach necessarily requires elevated excitation densities, which may in some cases alter the apparent exciton diffusion by inducing lattice heating or secondary nonradiative effects. For example, exciton-exciton scattering in $WS_2$ is reported to cause a non-linear increase in $D$ with increasing excitation power.[28] Furthermore, because the probability of exciton–exciton annihilation depends non-linearly on excitation density, experimental factors such as non-uniform (*e.g.*, gaussian) excitation profiles also complicate extraction of reliable quantitative results.

Here, we report the use of lanthanide impurity ions in $CrI_3$ as highly effective probes of exciton diffusion in this material. As luminescence activators, lanthanides harvest individual $CrI_3$ excitons *via* direct CrI$_3$ → Yb$^{3+}$ energy transfer (ET), eliminating the non-linear dependence of exciton-exciton annihilation on excitation density and allowing measurements to be performed at arbitrarily low excitation densities where lattice heating and other photo-induced artifacts are eliminated. Previous work on Yb$^{3+}$-doped CrX$_3$ compounds has revealed strong dopant-lattice magnetic superexchange coupling that pins Yb$^{3+}$ spins to those of the surrounding CrX$_3$ lattice.[29,30] These superexchange interactions rely on covalency, which also promotes CrX$_3$ → Yb$^{3+}$ ET through the well-known Dexter mechanism. Using a combination of continuous-wave (CW) and time-resolved PL measurements performed as a function of temperature and Yb$^{3+}$ doping level, complemented by Monte Carlo modeling, we develop a detailed description of exciton-diffusion dynamics in CrI$_3$. The results demonstrate that exciton diffusion is restricted by strong electron-nuclear coupling in the emissive $^4T_{2g}$ ($O_h$) excited state, with a low-temperature value of $D = 1.4 \times 10^{-8}$ cm$^2$/s that increases to $2.2 \times 10^{-7}$ cm$^2$/s by room temperature. Above ~100 K, the PL data show evidence for thermally activated nonradiative relaxation *via* phonon emission that ultimately limits exciton diffusion lengths to below ~3 nm at all temperatures, *i.e.*, well below the optical diffraction limit. These results thus provide a quantitative description of exciton diffusion in CrI$_3$ that will aid the interpretation of other photophysical phenomena such as exciton-exciton annihilation and, more broadly, that enriches our fundamental understanding of exciton physics in the CrX$_3$ family of strongly correlated magnetic materials.

**Results and Discussion**

**A. Experimental Results**

**General Considerations.** In idealized $O_h$ symmetry, the five Cr 3$d$ orbitals are split into $t_{2g}$ and $e_g$ sets by the crystal field, and the three unpaired spins in the degenerate $t_{2g}$ orbital yield a $^4A_{2g}$ ground state. The lowest-energy $^4T_{2g}$ ligand-field excited state corresponds to excitation of one $t_{2g}$ electron into a σ-antibonding $e_g$ orbital. In actuality, this formally parity-forbidden transition in CrI$_3$ gains electric-dipole allowedness through static symmetry lowering ($C_2$ site symmetry), boosted by configuration interaction with low-energy ligand-to-metal charge-transfer excited states.[31] Figure 1a depicts a single-configurational-coordinate diagram describing the $^4A_{2g}$ → $^4T_{2g}$ absorption and luminescence transitions for the $A_{1g}$ local nuclear coordinate. For simplicity, the comparably large displacement expected along the $E_g$ (Jahn-Teller) nuclear coordinate[32] is not shown. Excited-state distortion along these two nuclear coordinates ($\Delta Q_i$, $i$ = $A_{1g}$, $E_g$) stabilizes that state by a reorganization energy, $E_R$, as described by eq 1.

$$E_R = \sum_i \frac{1}{2} k_i (\Delta Q_i)^2 \qquad (1)$$

Spectroscopically, $E_R$ can be estimated from the PL Stokes shift through $E_{Stokes} = 2E_R$, giving $E_R \approx 200$ meV for CrI$_3$.[33] Importantly, when $E_R$ is so large, exciton hopping from one Cr$^{3+}$ site to a neighboring site also requires a large nuclear reorganization, as depicted schematically in Figure 1b, and this reduces the microscopic site-to-site ET rate constant, $k_{Cr}^{hop}$.

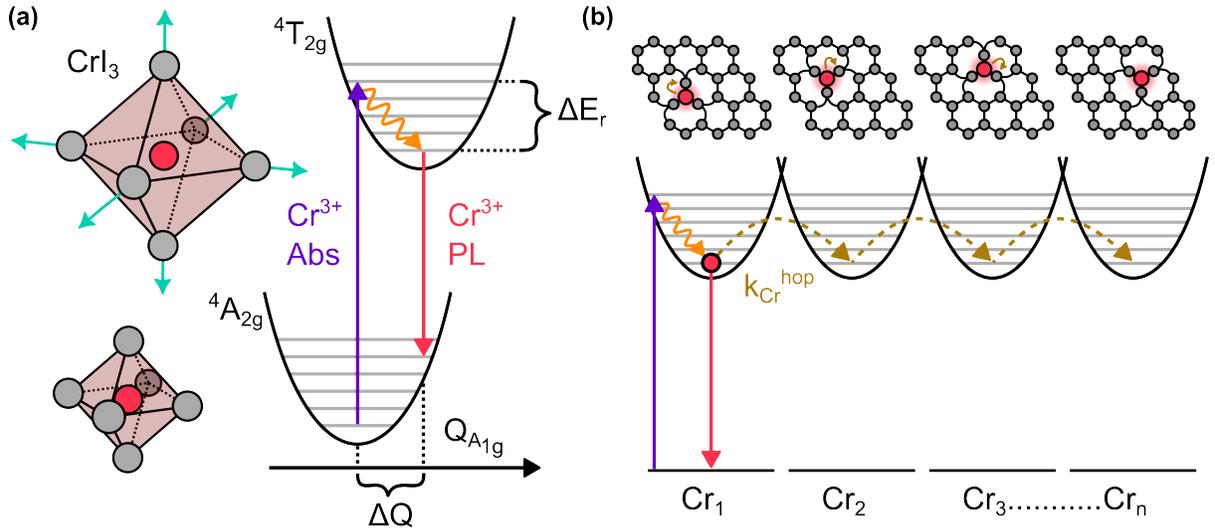

**Figure 1. (a)** Single-configurational-coordinate (SCC) diagram describing nuclear reorganization along the $A_{1g}$ coordinate during transitions between the $^4A_{2g}$ and $^4T_{2g}$ ligand-field states of octahedral Cr$^{3+}$ in idealized $O_h$ symmetry. The pictures illustrate the $A_{1g}$ distortion in the [CrI$_6$]$^{3-}$ excited state. Additional displacement along the $E_g$ nuclear coordinate (not shown) is also expected. The large nuclear reorganization energy ($E_R$) associated with these distortions stabilizes excitons at individual Cr$^{3+}$ sites. The actual Cr$^{3+}$ site symmetry is lower ($C_2$) and lacks inversion. **(b)** Schematic depiction of site-to-site exciton hopping involving short-range Dexter energy transfer between Cr$^{3+}$ ions.

**Time-Resolved Photoluminescence Spectroscopy of Yb$^{3+}$:CrI$_3$.** Figure 2a-d shows 4 K PL spectra of a series of $x$% Yb$^{3+}$:CrI$_3$ single crystals with $x$ = 0.0, 0.2, 1.6, and 18.6 (see Methods). With increasing Yb$^{3+}$ concentration there is a corresponding increase in the relative intensity of sharp-line Yb$^{3+}$ $^2F_{5/2} \rightarrow {}^2F_{7/2}$ PL centered ~ 1.135 eV, and a concomitant suppression of the broad Cr$^{3+}$ $^4T_{2g} \rightarrow {}^4A_{2g}$ PL centered ~ 1.08 eV. Figure 2e plots 4 K PL decay curves for the samples shown in Figure 2a-d, all collected at 1.20 eV and therefore corresponding solely to $Cr^{3+}$ PL. The $Cr^{3+}$ decay time shortens from ~0.8 μs in undoped $CrI_3$ to ~0.2 μs at the highest $Yb^{3+}$ concentration (18.6%), and the full data set is summarized in Figure 2e(inset). Despite a small amount of scatter, the trend of a decreasing $\tau_{Cr}$ with increasing concentration of $Yb^{3+}$ is clear. We speculate that the scatter comes from sample-to-sample variations in intrinsic defects. Figure 2f plots 4 K PL decay dynamics for the same samples but now collected at 1.15 eV, where both $Cr^{3+}$ and $Yb^{3+}$ emit. In addition to the same $CrI_3$ decay seen in Figure 2e, a new $Yb^{3+}$ signal is observed with a concentration-independent decay time of ~10 μs but whose relative intensity increases with increased $Yb^{3+}$ doping. These data demonstrate $CrI_3$ exciton annihilation by $Yb^{3+}$ dopants. Critically, the ability to independently probe both participants involved in the annihilation ($CrI_3$ and $Yb^{3+}$) offers a rare opportunity to gain additional insight into the underlying $CrI_3$ photophysics.

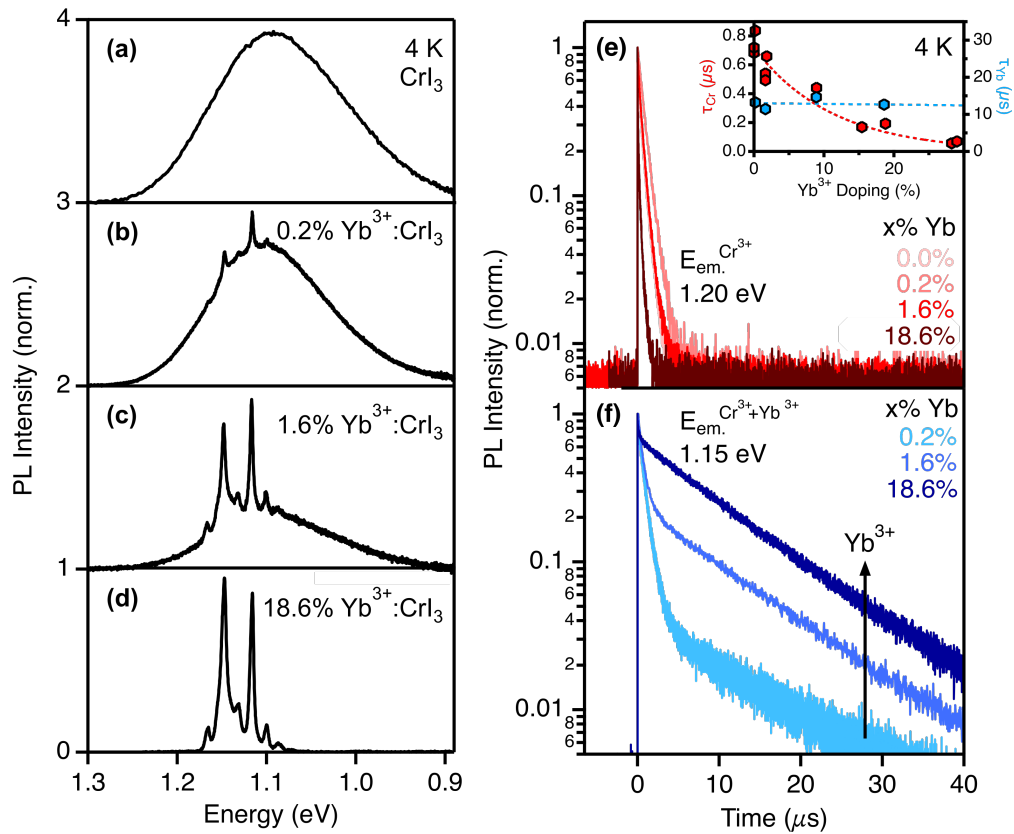

**Figure 2.** PL spectra of (**a**) 0.0% $Yb^{3+}$:$CrI_3$, (**b**) 0.2% $Yb^{3+}$:$CrI_3$, (**c**) 1.6% $Yb^{3+}$:$CrI_3$, and (**d**) 18.6% $Yb^{3+}$:$CrI_3$ measured at 4 K using CW excitation at 1.88 eV. (**e-f**) 4 K PL decay curves measured for the same samples using a 40 ps, 50 Hz pulsed laser excitation source at 532 nm (2.33 eV) and a 6.27 nm spectral bandwidth for detection. The sample PL was measured at two regions 1.20 eV (e) and 1.15 eV (f), representing CrI$_3$ emission (Cr$^{3+}$) and a superposition of CrI$_3$ and dopant emission (Yb$^{3+}$ + Cr$^{3+}$), respectively. Inset: Cr$^{3+}$ (red) and Yb$^{3+}$ (blue) PL decay times plotted vs Yb$^{3+}$ concentration, *x*. With increased *x*, the Cr$^{3+}$ decay time shortens while the Yb$^{3+}$ PL decay time remains relatively constant. The dashed lines are guides to the eye.

Figure 3a shows gated 4 K PL spectra of the 18.6% Yb$^{3+}$ sample collected by integrating only the first 50 ns following the excitation pulse ($t_{delay}$ = 25 ns, $\Delta t$ = 50 ns). The spectrum consists of ~8% Yb$^{3+}$ (integrated) intensity and 92% Cr$^{3+}$ (integrated) intensity. By 225 ns after the excitation pulse ($t_{delay}$ = 225 ns, $\Delta t$ = 50 ns), the spectrum has evolved to become 30% Yb$^{3+}$ and 70% Cr$^{3+}$. The CW spectrum shows >99% Yb$^{3+}$ PL. These data indicate that CrI$_3$ exciton capture by Yb$^{3+}$ occurs on the timescale of hundreds of nanoseconds. The two independent PL signals can thus be used to quantify the exciton-annihilation dynamics. To illustrate, Figure 3b,c plots 4 K PL decay traces collected at 1.20 eV (only Cr$^{3+}$ PL) and 1.15 eV (Cr$^{3+}$ and Yb$^{3+}$ PL) for the 0.2% and 1.6% Yb$^{3+}$ samples shown in Figure 2, focusing on the early times. Subtracting the former curves from the latter isolates the time evolution of just the Yb$^{3+}$ PL. These Yb$^{3+}$ curves are also plotted in Figure 3b,c, and both show a distinct rise in Yb$^{3+}$ PL following the laser pulse, reflecting the time required for CrI$_3$ excitons to be harvested by Yb$^{3+}$ impurities. Fitting these Yb$^{3+}$ curves to biexponential (rise + decay) functions yields rise times of $\tau_{rise}$ = 684 ± 1 ns for the 0.2% Yb$^{3+}$ sample and 268 ± 3 ns for the 1.6% Yb$^{3+}$ sample. These results demonstrate faster CrI$_3$ exciton annihilation in samples with more Yb$^{3+}$, as anticipated.

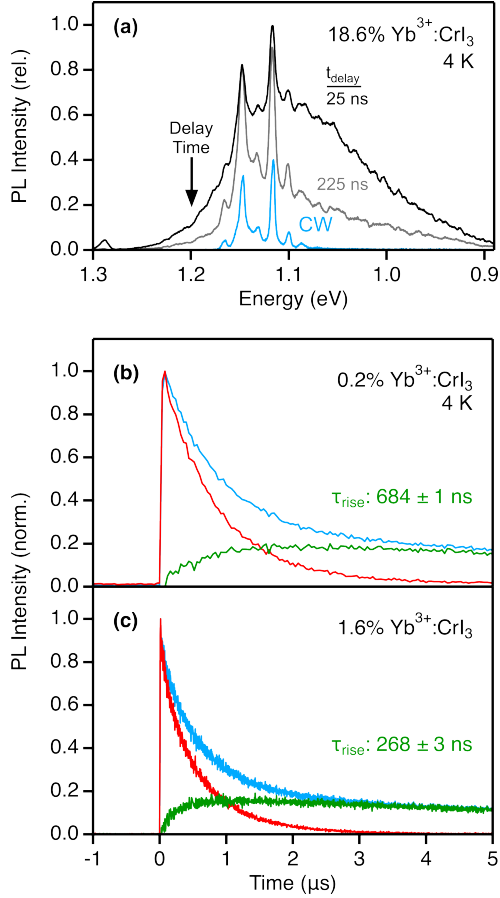

**Figure 3. (a)** PL spectrum of 18.6% $Yb^{3+}$:$CrI_3$ measured at 4 K using CW excitation and time-integrated detection (blue line), compared to gated PL spectra of the same sample collected at 4 K. The black and grey spectra correspond to center times of 25 and 225 ns, respectively, each with integration windows of 50 ns following a 40 ps laser pulse ($E_{ex}$ = 2.33 eV). **(b, c)** PL decay curves collected at 4 K for (b) 0.2% and (c) 1.6% $Yb^{3+}$:$CrI_3$ following 40 ps excitation at 2.33 eV. The red and blue decay traces were collected at 1.20 and 1.15 eV and correspond to host-lattice emission ($Cr^{3+}$) and a superposition of host and dopant emission ($Yb^{3+}$ + $Cr^{3+}$), respectively, shown in Figure 2e,f. The green curves plot the differences between these two decay traces normalized at $t = 0$ for each sample, showing rise times for $Yb^{3+}$ emission of $\tau_{rise}$ = 684 ± 1 ns and 268 ± 3 ns for 0.2 and 1.6% $Yb^{3+}$:$CrI_3$, respectively.

**Variable-Temperature Photoluminescence Spectroscopy of $Yb^{3+}$:$CrI_3$.** To explore the effect of temperature on exciton diffusion and annihilation, PL measurements were performed at several temperatures. Figure 4a-d shows variable-temperature PL (VTPL) spectra collected from 4 to 200 K for the series of $x$% $Yb^{3+}$:$CrI_3$ samples shown in Figure 2. All samples show a monotonic reduction in total integrated PL intensity with increasing temperature, and no PL above ~200 K. With increasing temperature, the $Yb^{3+}$ PL features at all doping levels broaden due to thermal spin disorder[29] and growth of hot bands. At all temperatures, the 0.2% and 1.6% $Yb^{3+}$ samples (Figure 4b,c) display a mixture of broad $Cr^{3+}$ emission and narrow $Yb^{3+}$ emission, whereas the 18.6% $Yb^{3+}$:$CrI_3$ (Figure 4d) displays only $Yb^{3+}$ emission.

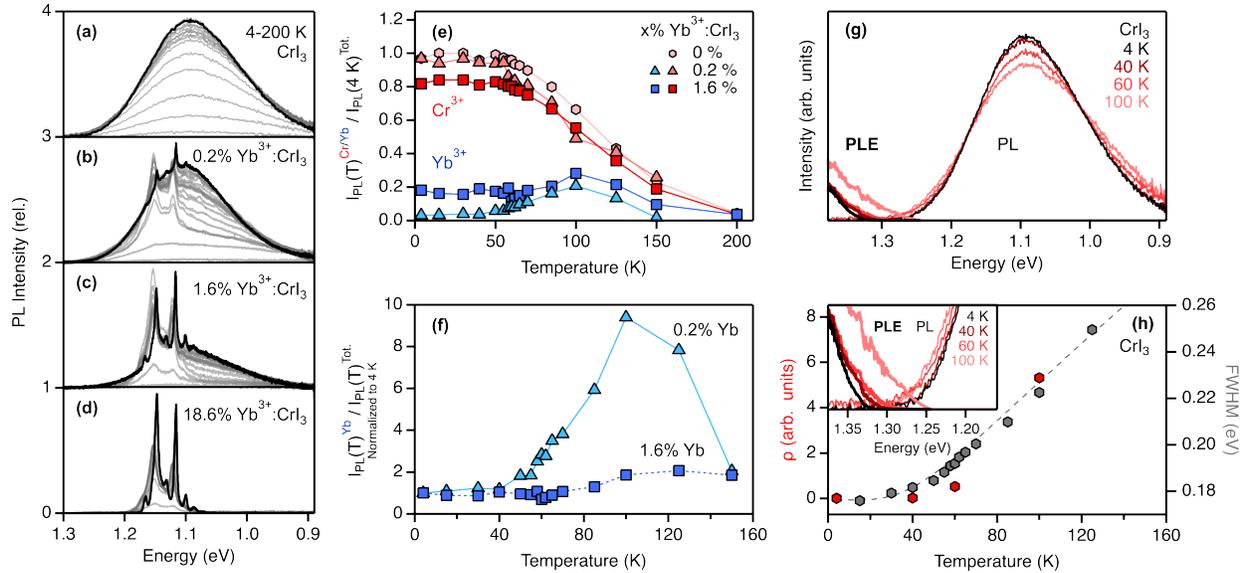

**Figure 4.** VTPL spectra of (**a**) 0.0% $Yb^{3+}$:$CrI_3$, (**b**) 0.2% $Yb^{3+}$:$CrI_3$, (**c**) 1.6% $Yb^{3+}$:$CrI_3$, and (**d**) 18.6% $Yb^{3+}$:$CrI_3$ measured from 4 to 200 K using CW excitation at 1.88 eV. (**e**) Plots of the integrated $Cr^{3+}$ (red) and $Yb^{3+}$ (blue) PL intensities divided by the total integrated PL intensity at 4 K for each sample in panels (a-d). See Figure S1 for further details. All samples show a similar temperature dependence. (**f**) Plot of integrated $Yb^{3+}$ as a portion of the total luminescence intensity normalized to their 4 K value for the 0.2% and 1.6% sample. (**g**) Plot of PLE (bold) and PL spectra for $CrI_3$ at 4, 40, 60, and 100 K. The PL spectra were normalized to a unit area. (**h**) Plot of the FWHM of $CrI_3$ (grey) overlayed with the calculated spectral overlap derived from panel (g). Inset: expanded plot of the spectra shown in panel (g) highlighting the region of spectral overlap.

Because $Yb^{3+}$ emission only occurs after a $CrI_3$ exciton is annihilated by a dopant, the integrated $Yb^{3+}$ PL intensity directly reflects the number of excitons that have diffusively encountered a dopant. Figure 4e plots the temperature dependence of the integrated $Cr^{3+}$ ($I_{PL}^{Cr}$) and $Yb^{3+}$ ($I_{PL}^{Yb}$) PL intensities divided by the total integrated PL intensity at 4 K ($I_{PL}^{tot}$(4K)). These ratios convey the PL fraction coming from each center ($Cr^{3+}$ or $Yb^{3+}$) relative to the total intensity measured at 4 K. The fractional $Cr^{3+}$ PL ($I_{PL}^{Cr}(T)/I_{PL}^{tot}(4K)$) is temperature-independent up to ~60 K and then steadily decreases with increasing temperature up to 200 K, independent of $Yb^{3+}$ doping. Conversely, in 0.2% $Yb^{3+}$:$CrI_3$, the ratio $I_{PL}^{Yb}(T)/I_{PL}^{tot}(4K)$ is temperature-independent from 4 to ~60 K before *rising* from ~0.03 to a maximum of 0.20 at 100 K, and then dropping at higher temperatures. 1.6% $Yb^{3+}$:$CrI_3$ shows a larger value of $I_{PL}^{Yb}(T)/I_{PL}^{tot}(4K)$ at 4 K (0.18), which rises to 0.28 at 100 K. Figure 4f recasts these data as $I_{PL}^{Yb}(T)/I_{PL}^{tot}(T)$ normalized to $I_{PL}^{Yb}(4K)/I_{PL}^{tot}(4K)$. In both samples, the $Yb^{3+}$ PL intensity remains constant from 4 K up to ~50 to 60 K and then increases with increasing temperature. Above ~100 K, the $Yb^{3+}$ PL intensity drops steeply to zero.

These data show exciton annihilation by $Yb^{3+}$ becoming more probable at higher temperature, indicating that $CrI_3$ exciton diffusion is thermally activated. This temperature dependence is distinct from that of $CrI_3$ spin correlation, however, as represented by the infinite spin-correlation function $(M(T)/M(0))^2$, where M is magnetization (Figure S4). This spin-correlation function decreases immediately starting at the lowest temperatures before dropping sharply to zero around $T_C$ = 61 K. The PL temperature dependence in Figure 4 thus shows no evidence of being affected by spin ordering, consistent with expectations for Dexter energy transfer. This insensitivity to spin differs from that observed in AFM-ordered $(CH_3)_4NMnCl_3$, where exciton migration is reduced upon magnetization due to suppression of dipole-dipole $Mn^{2+}$-$Mn^{2+}$ energy transfer, which requires a change of spin state ($^4T_1$ to $^6A_1$) on each ion.[34] We also observe no spin contribution to exciton diffusion in time-resolved PL measurements around $T_C$ (Figure S5). Instead, we attribute the PL temperature dependence to thermal spectral broadening.

**Temperature-Dependent Exciton Hopping.** Strong electron-phonon coupling in the luminescent excited state of $CrI_3$ effectively confines an exciton to a single $Cr^{3+}$ site. This exciton may still diffuse by incoherent hopping between individual lattice sites (Figure 1b),[35] with a temperature-dependent hopping rate constant ($k_{Cr}^{hop}(T)$) that can be described within the framework of Fermi's golden rule, as shown in eq 2.

$$k_{Cr}^{hop}(T) = \frac{2\pi}{\hbar}|M_{DA}|^2\rho(T) \qquad (2)$$

Here, $M_{DA}$ describes the donor-acceptor electronic coupling and $\rho(T)$ is the temperature-dependent spectral-overlap factor, defined in eq 3, where $\bar{f}_D(E)$ represents the area-normalized donor emission spectrum and $\bar{\varepsilon}_A(E)$ represents the area-normalized absorption spectrum of the acceptor, obtained experimentally by PL excitation (PLE).

$$\rho = \int_0^\infty \bar{f}_D(E)\bar{\varepsilon}_A(E)dE \qquad (3)$$

Figure 4g plots PLE spectra of undoped $CrI_3$ measured at 4, 40, 60, and 100 K, collected by monitoring the $Cr^{3+}$ emission at 1.0 eV and normalizing by the relative PL intensity at each temperature. The corresponding PL spectra are also shown for each temperature, also normalized to relative PL intensity The PL spectrum is unchanged from 4 to ~50 K, but the band broadens noticeably at higher temperatures. A similar trend is observed in the PLE spectrum. The inset to Figure 4h plots the overlap region of Figure 4g on an expanded scale, showing that $\rho$ increases with increasing temperature due to this spectral broadening. Using eq 3, the data in Figure 4g can be analyzed to quantify the temperature dependence of $\rho$, and the results are plotted in red in Figure 4h. Figure 4h also plots values for the full widths at half-maximum (FHWM) determined from the PL spectra of undoped $CrI_3$ measured between 4 and 160 K (Figure 4a). The FHWM and $\rho$ values show very similar trends with temperature, both remaining constant at low temperatures before increasing above ~50 K. With increasing temperature, higher vibrational excited states in the $^4A_{2g}$ ground and $^4T_{2g}$ excited electronic states of $Cr^{3+}$ are thermally populated, leading to new absorption and PL intensities that increase the spectral overlap ($\rho$). More quantitatively, the FWHM data were fit to the second-moment hyperbolic cotangent function of eq 4,[36] where $v_{eff}$ represents the energy of an effective activating vibration. The best-fit value of $v_{eff} = 97$ cm$^{-1}$ is close to the energies of the dominant vibrational modes observed in $CrI_3$ resonance Raman spectra[37] and consistent with expectations for the $A_{1g}$ and $E_g$ local modes of $[CrI_6]^{3-}$.

$$FWHM(T) = FWHM(0)\left[\coth\left(\frac{v_{eff}}{2k_B T}\right)\right]^{1/2} \qquad (4)$$

Importantly, the temperature dependence of $\rho$ is very similar to those of the $Cr^{3+}$ and $Yb^{3+}$ PL intensities (Figure 4e,f), confirming the importance of thermal spectral broadening in $CrI_3$ exciton diffusion and annihilation by $Yb^{3+}$ and thereby confirming the critical role that nuclear reorganization plays in limiting exciton diffusion in $CrI_3$.

Exciton diffusion in $CrI_3$ can also be probed by monitoring the dependence of $Cr^{3+}$ PL decay dynamics on $Yb^{3+}$ doping concentration at various temperatures. Figure 5a summarizes the temperature dependence of the $Cr^{3+}$ PL decay times ($\tau_{Cr}$) for a series of $x\%$ $Yb^{3+}$:$CrI_3$ samples ($x$ = 0.0, 0.2, 1.6, 8.9, 18.6). At a fixed temperature, increasing $x$ shortens $\tau_{Cr}$, *e.g.*, from ~0.80 ms in 0.2% $Yb^{3+}$ to only ~0.15 ms in 18% $Yb^{3+}$ at 4 K. For all samples, $\tau_{Cr}$ is independent of temperature up to ~50 K, above which it decreases with increasing temperature. When these $\tau_{Cr}$ data are normalized to their low-temperature values as shown in Figure 5b, it is evident that increased $Yb^{3+}$ doping also increases the temperature dependence of $\tau_{Cr}$: the higher-doped samples show a steeper decrease in $\tau_{Cr}$ above 50 K.

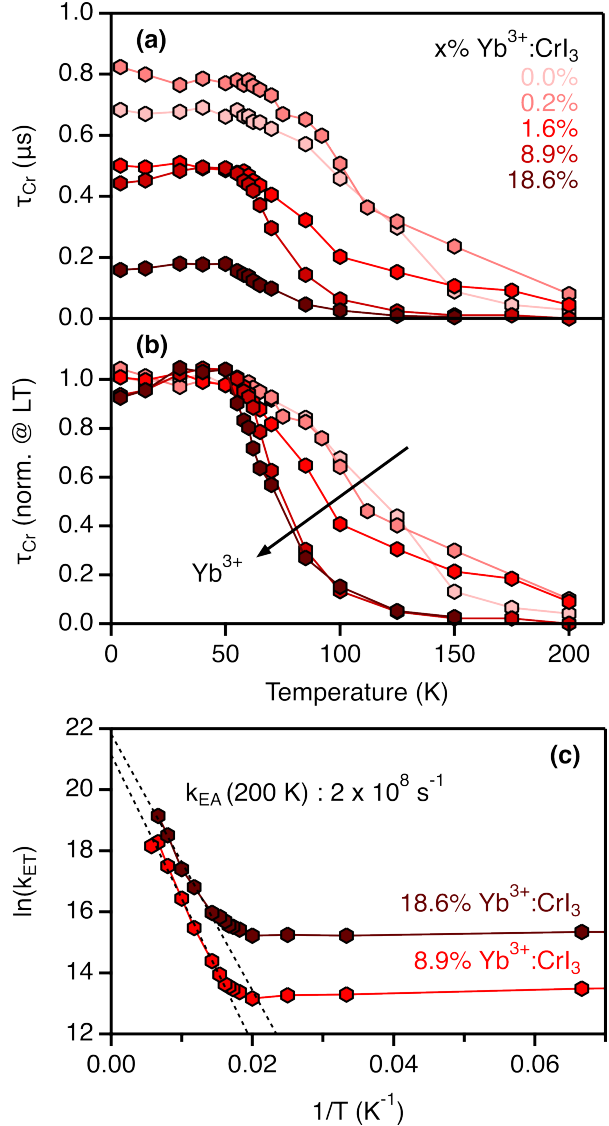

**Figure 5. (a)** $CrI_3$ PL decay times plotted vs temperature for $x$% $Yb^{3+}:CrI_3$ ($x$ = 0.0, 0.2, 1.6, 8.9, 18.6), measured following 40 ps excitation at 2.33 eV. Increased $x$ shortens the $CrI_3$ decay time. The shorter decay time of the undoped $CrI_3$ compared to the 0.2% $Yb^{3+}:CrI_3$ sample is attributed to its reduced crystal quality. **(b)** The data from panel (a) normalized to the lowest temperature region. Increasing $x$ increases the rate of PL quenching with temperature. **(c)** Plot of the effective energy-transfer rate constant ($k_{ET}$) vs temperature for 8.9 and 18.6% $Yb^{3+}:CrI_3$ samples. The dashed lines show best fits of the data to the Arrhenius model of eq

6.

The data in Figure 5a can now be used to estimate effective exciton annihilation rate constants ($k_{EA}$) following the approach of Kambli et al.,[38] summarized in eq 5.

$$k_{EA} = k_{Cr\_doped} - k_{Cr\_undoped} = 1/\tau_{Cr\_doped} - 1/\tau_{Cr\_undoped} \qquad (5)$$

$k_{EA}$ consists of two processes: (*i*) the multiple $Cr^{3+} \rightarrow Cr^{3+}$ hopping steps that occur as the exciton migrates through the lattice prior to capture by $Yb^{3+}$, individually described microscopically by $k_{Cr}^{hop}$, and (*ii*) the final $Cr^{3+} \rightarrow Yb^{3+}$ energy-transfer step, described by $k_{Yb}^{hop}(T)$. Because $k_{Yb}^{hop} > k_{Cr}^{hop}$ at all temperatures (*vide infra*), $k_{Yb}^{hop}$ is effectively temperature independent in these measurements. Consequently, the experimental data in Figure 5a,b directly reflect the temperature dependence of $k_{Cr}^{hop}$. At low temperatures, $Cr^{3+} \rightarrow Cr^{3+}$ hopping is slow and thus $\tau_{Cr}$ is at its maximum. In this regime, increasing the concentration of $Yb^{3+}$ reduces the number of hops required before exciton annihilation by $Yb^{3+}$, as shown by the decrease in $\tau_{Cr}$. Above ~50 K, $k_{Cr}^{hop}$ increases, and this accelerates exciton annihilation in all samples.

Figure 5c presents the 8.9 and 18.6% $Yb^{3+}$ $\tau_{Cr}$ data from Figure 5a as $k_{EA}$ values (converted using eq 5) plotted in Arrhenius form, highlighting the thermal activation of exciton diffusion in $CrI_3$. Fitting the high-temperature slope to eq 6 yields $k_{EA}(\infty) \approx 2 \times 10^9$ s$^{-1}$, representing the hypothetical scenario in which $k_{Cr}^{hop}$ is still rate limiting at infinite temperature. The fact that $k_{EA}$ is always larger in 18.6% $Yb^{3+}$ than in 8.9% $Yb^{3+}$ indeed indicates that $k_{Cr}^{hop}$ is rate determining at all temperatures probed experimentally. $k_{EA}$ at our highest experimental temperature (200 K) thus provides an experimental lower bound of $2 \times 10^8$ s$^{-1}$ for the value of $k_{Yb}^{hop}$. This value is comparable to, for example, the lower limit of $k_{EA} \sim 10^8$ s$^{-1}$ reported for energy transfer in structurally well-defined cyanide-bridged $Cr^{3+}$-$Yb^{3+}$ dimers at 300 K.[39]

$$k_{EA}(T) = k_{EA}(\infty)e^{-\Delta/k_BT} \qquad (6)$$

When considered together with the preceding experimental data, it is clear that exciton diffusion in $CrI_3$ is slow, thermally activated, and governed by nearest-neighbor hopping kinetics, and that the final ET step to $Yb^{3+}$ is sufficiently fast to never be rate-limiting.

**B. Kinetic Monte Carlo Simulations**

**General Considerations.** To describe the real-space and -time profiles of exciton diffusion in $CrI_3$, kinetic Monte Carlo (KMC) simulations were performed using parameters defined by the above experiments. Exciton diffusion was modeled as sequential nearest-neighbor random hopping events within hexagonal $CrI_3$ lattices containing randomly distributed $Yb^{3+}$ impurities at various designated concentrations. For $k_{Cr}^{hop}$, two possible mechanisms of energy transfer may be considered: through-space multipolar Förster-type resonant energy transfer (FRET) and through-bond Dexter-type energy transfer (DET), both of which may be present. Previous work has shown that ferromagnetic order in $CrI_3$ arises from high Cr-I covalency, which enhances the ferromagnetic superexchange coupling.[40-43] Because of this high covalency, intra-layer DET is expected to dominate $k_{Cr}^{hop}$, an assumption validated below. Figure 6a illustrates the full set of rate constants governing specific state transitions applied in this work.

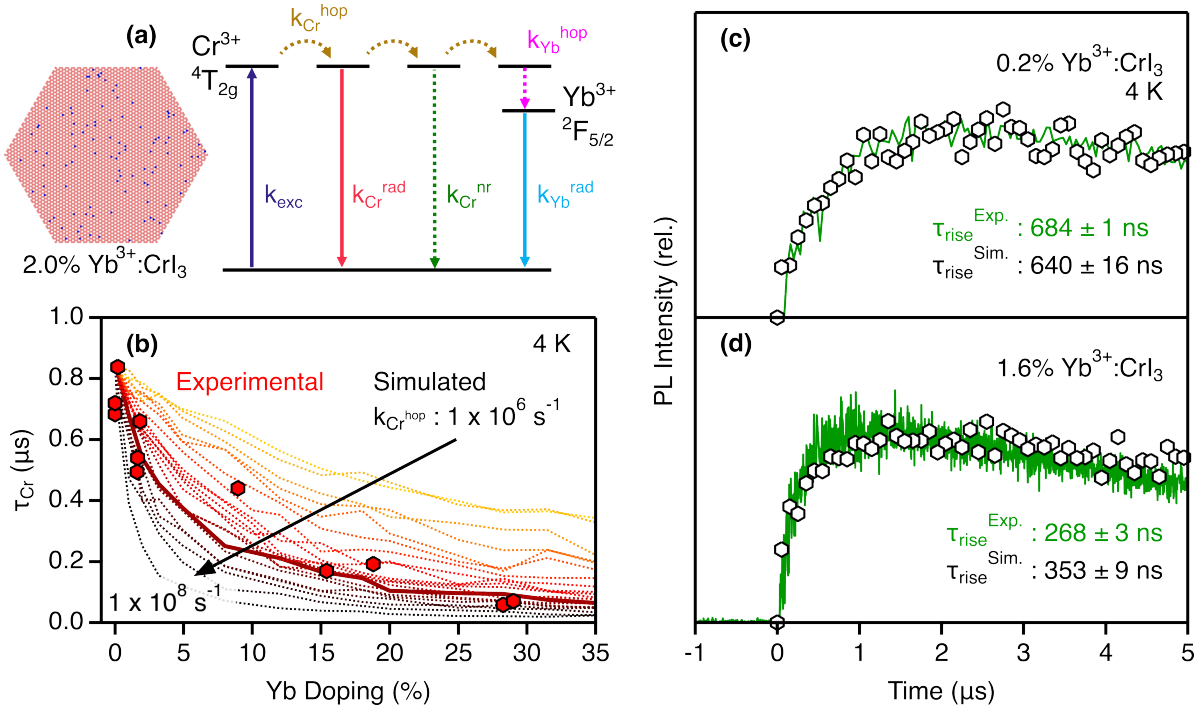

**Figure 6. (a)** A representative trial crystal with random $Yb^{3+}$ doping at 2% of the cation sites. Each red (blue) pixel corresponds to an octahedrally coordinated $Cr^{3+}$ ($Yb^{3+}$) ion. The energy diagram illustrates the processes and rate constants addressed in the Monte Carlo simulation. Values $k_{Cr}^{rad}$, $k_{Cr}^{nr}$, $k_{Yb}^{rad}$, $k_{Yb}^{hop}$ were all determined experimentally (see main text), while $k_{Cr}^{hop}$ was determined *via* simulation as shown in panel (b). These processes compete with one another on each $Cr^{3+}$ ion but are illustrated on different ions for ease of viewing. **(b)** Plot of experimental $Cr^{3+}$ decay times at 4 K for various $Yb^{3+}$ doping levels (red scatter points). The curves show simulated values of the 4 K $Cr^{3+}$ decay times calculated over many values of $k_{Cr}^{hop}$. The bold red line corresponds to the best fit, yielding a value of $k_{Cr}^{hop}$ = 1 x 10$^7$ s$^{-1}$ under these conditions. **(c-d)** Plot of simulated Yb$^{3+}$ PL dynamics (white scatter points) for (c) 0.2% and (d) 1.6% Yb$^{3+}$:CrI$_3$ at 4 K, calculated using $k_{Cr}^{hop}$ = 1.0 x 10$^7$ s$^{-1}$. The corresponding experimental data (green, from Figure 3b,c) are included for comparison.

Simulations were performed by constructing trial crystals of Yb$^{3+}$-doped CrI$_3$. Each crystal contained 5100 M$^{3+}$ octahedra ordered into the hexagonal structure of monolayer CrI$_3$. For each Yb$^{3+}$ doping level, 250 unique crystals were generated, each with a random distribution of the appropriate number of Yb$^{3+}$ dopants. One such trial crystal with 2% Yb$^{3+}$ doping is shown in Figure 6a. For each of these trial crystals, 250 excitons were simulated, and all results were aggregated across the 250 crystals, resulting in the simulation of a total of 62,500 excitons for each Yb$^{3+}$ doping level at each temperature. Each simulation was launched with an exciton at the central ion of the crystal, and the fate of this exciton was tracked over time as it migrated around the lattice stochastically and eventually decayed, as determined by the given rate constants. Exciton-exciton annihilation was not considered in the simulation because of the very low excitation densities used experimentally (2 nJ/cm$^2$, or one exciton per ~7 × 10$^6$ Cr$^{3+}$ ions, see SI). The time of emission, the identity of the emitting site (Cr$^{3+}$ or Yb$^{3+}$), and the position in space of the emission event were all recorded for each excitation. See Methods for further details about the KMC simulations.

**Experimental Definition of KMC Parameters.** To constrain the KMC simulations, several rate constants were defined based on the experimental results presented above, as described here: The observed rate constant for CrI$_3$ excited-state decay has radiative ($k_{Cr}^{rad}$) and nonradiative ($k_{Cr}^{nr}$) contributions, as described by eq 7.

$$\frac{1}{\tau_{Cr}} = k_{Cr} = k_{Cr}^{rad} + k_{Cr}^{nr} \tag{7}$$

As shown in Figure S7, the PL intensity ($I_{PL}$) and $\tau_{Cr}$ for undoped CrI$_3$ show identical temperature dependence, indicating that the decrease in each with increasing temperature stems from thermally activated nonradiative decay and that $k_{Cr}^{rad}$ is effectively independent of temperature, as expected for Cr$^{3+}$ with $C_2$ site symmetry. Substituting a thermally activated Mott-Seitz-type expression for $k_{Cr}^{nr}$ as shown in eq 8,

$$k_{Cr}^{nr}(T) = k_{Cr}^{nr}(0) \cdot \exp(-E/k_B T) \tag{8}$$

followed by rearrangement, gives eq 9, which describes the temperature-dependent excited-state lifetime.

$$\tau_{Cr}(T) = \frac{1}{k_{Cr}^{rad} + (k_{Cr}^{nr}(0) \cdot \exp(-E/k_B T))} \tag{9}$$

Fitting the $\tau_{Cr}(T)$ data for CrI$_3$ (Figure S7) yields $k_{Cr}^{nr}$(4 K) ~ 0 s$^{-1}$ increasing to 9.9 x 10$^6$ s$^{-1}$ at 200 K. A value of $k_{Cr}^{rad}$ = 1.9 x10$^6$ s$^{-1}$ was taken from the experimental $\tau_{Cr}$(4 K) of 0.2% Yb$^{3+}$:CrI$_3$, which was the longest decay time observed in our sample set. Similarly, $k_{Yb}^{rad}$ is taken as $1/\tau_{Yb}$ measured for the same sample at 4 K, where nonradiative decay is suppressed, giving a value of $k_{Yb}^{rad}$ = 7.1 x10$^4$ s$^{-1}$ (see Methods).

Doping CrI$_3$ with Yb$^{3+}$ makes the exciton decay dynamics sensitive to diffusion. In addition to $k_{Cr}^{hop}$, $k_{Yb}^{hop}$ is introduced to account for Cr$^{3+}$ → Yb$^{3+}$ energy transfer. For the simulations, $k_{Yb}^{hop}$ was set to a temperature-independent value of 2.0 x 10$^8$ s$^{-1}$ based on the high-temperature lower limit determined experimentally (Figure 5c), although variations in the precise value of this parameter have no effect because it is never rate limiting (see Figure 5 and discussion). To determine $k_{Cr}^{hop}(T)$, the low-temperature experimental data from Figure 2e are considered. The radiative and nonradiative decay constants found for CrI$_3$ ($k_{Cr}^{rad}$, $k_{Cr}^{nr}$) are intrinsic and hence treated as independent of Yb$^{3+}$ doping. Thus, only $k_{Cr}^{hop}(T)$ remains undefined by experiment. The full set of experimental parameters used in the KMC simulations is provided in Table S1.

**KMC Simulations to Determine $k_{Cr}^{hop}$.** Figure 6b shows simulated Cr$^{3+}$ decay times plotted *vs* temperature, calculated for the full range of Yb$^{3+}$ concentrations using the above fixed values for $k_{Cr}^{rad}$, $k_{Cr}^{nr}$, $k_{Yb}^{hop}$, with a range of values for $k_{Cr}^{hop}$ spanning over two orders of magnitude. The experimental data are best reproduced using $k_{Cr}^{hop}$(4 K) ≈ 1.0 x 10$^7$ s$^{-1}$ (Figure 6b, bold). This hopping rate constant is similar to those determined for Frenkel exciton migration in other ionic lattices such as RbMnF$_3$, CsMnF$_3$, and GdCl$_3$,[44-46] and critically, it is ~10x slower than the experimental lower limit of $k_{Yb}^{hop}$ from Figure 5. At the same time, this $k_{Cr}^{hop}$(4 K) value is ≥10$^3$x larger than the calculated 4 K rate constants for intra- and inter-layer nearest-neighbor FRET (Figure S11), supporting the assertion that exciton diffusion is limited to DET within the same monolayer.

As an independent cross-check, this value for $k_{Cr}^{hop}$ was used to simulate the experimental 4 K Yb$^{3+}$ PL dynamics of 0.2% and 1.6% Yb$^{3+}$:CrI$_3$ (Figure 3b,c). The simulated data are plotted in Figure 6c,d and reproduce the experimental results well, showing effective rise times of 640 ± 16 and 353 ± 9 ns, respectively. This comparison shows that the set of microscopic parameters used to simulate the 4 K Cr$^{3+}$ PL dynamics in Figure 6b also faithfully reproduces the 4 K Yb$^{3+}$ PL dynamics.

To simulate the full temperature range probed experimentally, the temperature dependence of $k_{Cr}^{hop}$ was analyzed. The temperature dependence of the PL FWHM can be expressed based on eq 4 as shown in eq 10.

$$\Delta FWHM = FWHM(T) - FWHM(0) = FWHM(0)\left[\coth\left(\frac{v_{eff}}{2k_BT}\right)\right]^{\frac{1}{2}} - FWHM(0) \quad (10)$$

From Figure 4h, $\rho$ shows the same temperature dependence as the FWHM, and from eq 2, $k_{Cr}^{hop}$ is proportional to $\rho$. Introducing a proportionality factor ($C$) into eq 10 thus allows substitution of $k_{Cr}^{hop}$ for FWHM, and rearrangement then gives eq 11.

$$\Delta k_{Cr}^{hop} = k_{Cr}^{hop}(T) - k_{Cr}^{hop}(0) = C\left[\left[\coth\left(\frac{v_{eff}}{2k_BT}\right)\right]^{\frac{1}{2}} - 1\right] \quad (11)$$

Following eq 11, experimental data were simulated across all temperatures and $Yb^{3+}$ concentrations using only a single universal adjustable scaling parameter ($C$), and the results are summarized in Figure 7a-c for the value of $C$ that best reproduces the experimental data. The best-fit value of $C = 8 \times 10^7$ s$^{-1}$ corresponds to a ~7-fold increase in $k_{Cr}^{hop}$ from $1 \times 10^7$ s$^{-1}$ at 4 K to $6.8 \times 10^7$ s$^{-1}$ at 200 K. The influence of $C$ on $k_{Cr}^{hop}$ is illustrated in Figure S10. Using this result, Figure 7a plots simulated $Cr^{3+}$ PL decay times at temperatures from 4 to 200 K for $x$% $Yb^{3+}$:CrI$_3$ ($x$ = 0.0, 0.2, 1.6, 5.0, 10.0, 18.6). These simulated results compare well with the corresponding experimental data in Figure 5a, covering a very broad range of compositions and temperatures. Overall, the KMC simulations reproduce all experimental variable-temperature, variable doping level, and time-dependence data well using a single experimentally constrained set of parameters, lending credence to the remaining parameters ($k_{Cr}^{hop}$ and $C$) determined by fitting to the data.

**Exciton Diffusion.** With these results, we are now able to analyze exciton diffusion in CrI$_3$ quantitatively. Figure 7b plots the exciton diffusion lengths ($L_D$) determined from the simulated trial data. $L_D$ is calculated as the average distance of exciton diffusion according to eq 12,

$$L_D = \sqrt{1/N \sum_i x_i^2} \quad (12)$$

where $N$ is the total number of excitons in a trial set and $x_i$ is the real-space difference between the initial and final positions of the $i$-th trial exciton. At 4 K, $L_D$ in undoped CrI$_3$ is 18.4 Å, corresponding to only ~5 lattice sites from the initial excitation. As the temperature increases, faster exciton hopping increases $L_D$, which reaches a maximum value of ~27 Å in undoped CrI$_3$ at ~100 K. At even higher temperatures, $L_D$ decreases again due to thermally activated nonradiative decay. $L_D$ also decreases with increasing $Yb^{3+}$ concentration at every temperature, reflecting the increasing probability of exciton annihilation by $Yb^{3+}$. At higher doping levels, $L_D$ becomes less sensitive to temperature because excitons are always efficiently annihilated after just a few hops, even at 4K.

Figure 7c plots exciton diffusion coefficients (diffusivities) determined at each temperature for each sample following eq 13.

$$D = \frac{L_D^2}{\tau} \quad (13)$$

In undoped $CrI_3$, $D$ increases from 1.4 x $10^{-8}$ $cm^2$/s at 4 K to 1.5 x $10^{-7}$ $cm^2$/s at 200 K. The temperature dependence of $D$ is linear above ~50 K, allowing facile extrapolation to a room-temperature value of $D$ = 2.2 x $10^{-7}$ $cm^2$/s. $D$ is also relatively insensitive to $Yb^{3+}$ doping level at all temperatures.

From these results, it is clear that exciton diffusivity in $CrI_3$ is several orders of magnitude lower than reported for other 2D materials such as $MoSe_2$, $WSe_2$, and the 2D perovskite $PEA_2PbI_4$, which all show $D \sim 10^{-1} - 10^1$ $cm^2$/s at 300 K, as probed by spatially resolved PL spectroscopy.[13, 47-49] As semiconductors, those materials feature far more delocalized excitons with much smaller nuclear reorganization energies, $E_R$. The low exciton diffusivity of $CrI_3$ is more similar to the exciton diffusivities reported for other ionic insulators,[44-46] for close-packed films of colloidal CdSe/CdS quantum dots,[50] and for organic materials,[51] all of which involve discrete site-to-site exciton hopping.

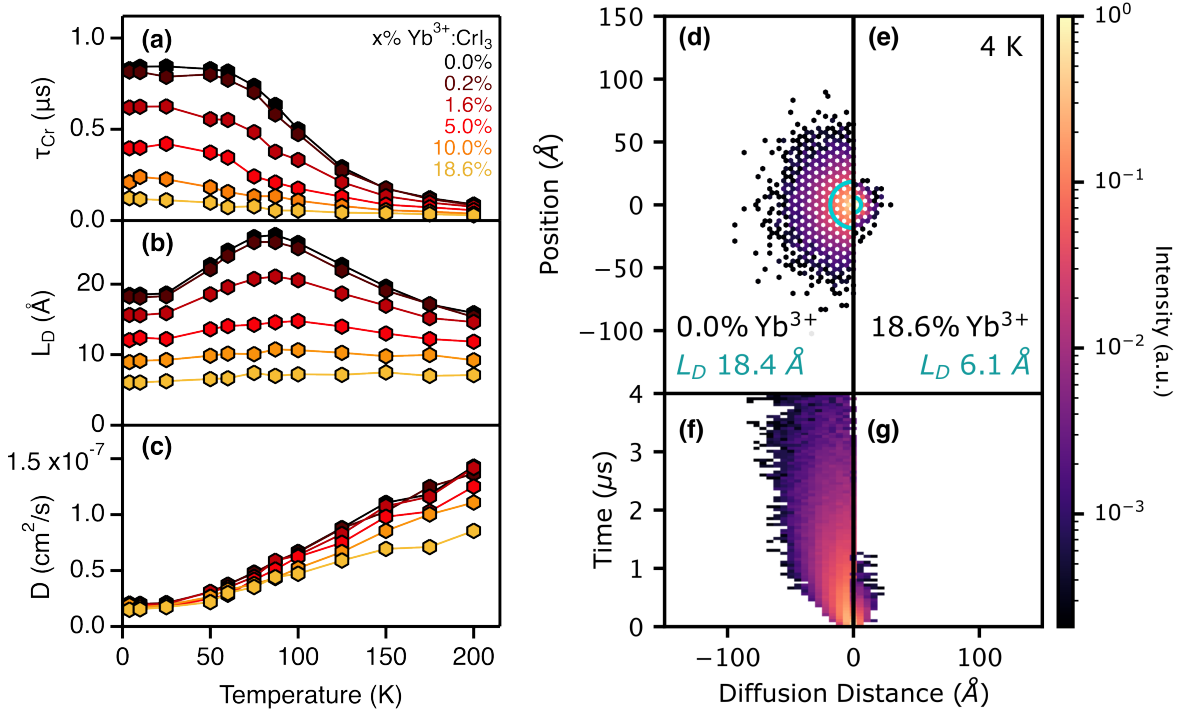

**Figure 7. (a)** Simulated $Cr^{3+}$ PL decay times obtained from Monte Carlo modeling for $x$% $Yb^{3+}$:$CrI_3$ ($x$ = 0.0, 0.2, 1.6, 5.0, 10.0, 18.6). **(b)** Simulated exciton diffusion lengths, $L_D$, plotted as a function of temperature. **(c)** Diffusion constant $D$ as a function of temperature, calculated as $D = L_D^2/\tau$. **(d-e)** False-color plots of simulated spatial PL mapping for 62,500 excitons in $CrI_3$ (d) and 18.6% $Yb^{3+}$:$CrI_3$ (e) at 4 K. The teal circle indicates $L_D$ for each simulation. All excitations began in the center of the crystal. **(f-g)** False-color plots of emission intensity as it evolves in space and time for the same data from panels (d-e).

To highlight the spatial and temporal aspects of these results, Figure 7d,e shows false-color simulated spatial PL maps for undoped $CrI_3$ and 18.6% $Yb^{3+}$-doped $CrI_3$ at 4 K, plotting the final location of each $CrI_3$ exciton relative to its initial position. Both profiles are gaussian, with the majority of PL occurring within the first few cation shells. The teal radii denote $L_D$ in each case. $L_D$ = 18.4 Å in undoped $CrI_3$ at 4 K, with some excitons traveling up to 100 Å, and this value decreases to 6.1 Å upon doping with 18.6% $Yb^{3+}$. Figure 7f,g plots the calculated PL intensities for these two samples as they evolve in space and time. Compared to the longer-lived excitons of undoped $CrI_3$, all excitons in 18.6 % $Yb^{3+}$-doped $CrI_3$ are quickly annihilated by nearby $Yb^{3+}$ ions. Notably, all $CrI_3$ samples have $L_D$ well below the optical diffraction limit at all temperatures.

**Conclusion**

We have demonstrated the use of emissive lanthanide impurities as atomic probes of exciton diffusion in $CrI_3$. These energy-harvesting dopants act as isolated atomic lattice sites for exciton annihilation, allowing exciton diffusion within the lattice to be systematically probed. We find that the highly localized excitons of $CrI_3$ diffuse through the lattice slowly *via* thermally activated site-to-site hopping driven by nearest neighbor Dexter energy transfer. Through a combination of time-resolved and continuous-wave PL spectroscopies supported by Monte Carlo simulations, we showed that the large nuclear reorganization reduces spectral overlap and impedes hopping, leading to exciton diffusivities of $10^{-7}$-$10^{-8}$ cm$^2$/s that are orders of magnitude lower than in 2D vdW semiconductors. Thus, exciton diffusion lengths in $CrI_3$ are also small, limited to < 3 nm at all temperatures, well below the optical diffraction limit. This understanding of $CrI_3$ exciton diffusivity should help to inform future studies of $CrI_3$ and related $CrX_3$ materials that are under investigation as spin-photonic components in advanced optoelectronics architectures.

More generally, as well as affording the opportunity to probe these otherwise inaccessibly small length scales of exciton diffusion, the use of impurity ions as exciton annihilators allows energy transport to be monitored at arbitrarily low excitation densities. Employment of lanthanide probes is not limited to compounds featuring slow exciton diffusion, such as $CrI_3$, but is extendable to a variety of interesting and complex compositions, including those involving heterointerfaces, which may itself offer interesting possibilities for investigating energy transport in novel van der Waals constructs. For example, it could be intriguing to study exciton diffusion in inorganic 2D perovskites by an analogous approach, because these materials feature self-trapped excitons and are increasingly being integrated into van der Waals heterostructure devices.[52-54] A particularly interesting material for such studies could be the 2D magnetic vdW compound $NiPS_3$.[55, 56] Recently, resonant inelastic x-ray scattering (RIXS) data with high energy resolution have shown evidence of dispersive excitons in $NiPS_3$ based on localized $Ni^{2+}$ $^3A_{2g} \rightarrow {}^1A_{1g}$ spin-flip excitations.[56] These excitons are proposed to propagate *via* a spin-conserving site-to-site hopping mechanism. Although not quantified, the weak electron-nuclear coupling in this $Ni^{2+}$ spin-flip excitation compared to the $Cr^{3+}$ $^4A_{2g} \rightarrow {}^4T_{2g}$ excitation of $CrI_3$ should enable greater diffusion lengths in the former, a hypothesis testable by the doping approach described here. Ultimately, any system in which impurity luminescence activators can be incorporated statistically at controlled and tunable concentrations offers the opportunity to probe exciton diffusion by the approach described here.

Whereas highly mobile excitons with large diffusion lengths are already amenable to study by direct PL profiling, the use of impurity luminescence activators excels when diffusion lengths are too small for such methods. Luminescent impurities, and especially lanthanides with their distinctively narrow line shapes, thus offer a powerful complementary tool for exploring energy transport in a diverse array of materials, paving the way for new insights and innovations in 2D optoelectronics.

**Methods**

**General Considerations.** Sample preparation and manipulation were conducted within a glovebox environment under a purified dinitrogen atmosphere.

**Chemicals.** For the chromium source, a high-purity chromium chip (99.995%, lot MKCH4484) purchased from Sigma Aldrich was ground into a fine powder using a mortar and pestle. Iodine (≥99.99%) was purchased from Sigma Aldrich, and ytterbium metal powder 40 mesh (99.9%) was obtained from BeanTown Chemical. All chemicals were used in their as-received state without further purification.

**Synthesis of $CrI_3$ and $Yb^{3+}$-Doped $CrI_3$ Single Crystals.** Doped and undoped $CrI_3$ single crystals were grown by chemical vapor transport as described previously.[29] Undoped $CrI_3$ was made by placing Cr(0) metal and $I_2$ at a 1:3 stochiometric ratio (per I atom) in a 15 cm long quartz tube with an inner diameter of 14 mm. The tubes were evacuated and flame sealed. The reaction tubes were positioned within an open-ended horizontal tube furnace, with the precursors situated at the center of the furnace set at 650 ˚C and the other end of the tube near the edge of the furnace at a temperature of ca. 500 ˚C. Samples were heated for 5 days and then allowed to gradually cool down to room temperature, yielding thin shiny black crystal flakes. For doped samples, additional Yb metal was added alongside the other reactants and the same procedure followed.

**Elemental Analysis.** Elemental analysis of the doped samples was performed by inductively coupled plasma mass spectrometry (ICP-MS) using a PerkinElmer NexION 2000B. Samples were digested in concentrated high-purity nitric acid with the aid of sonication, followed by dilution in ultrapure $H_2O$. $Yb^{3+}$ doping levels are reported as cation mole fraction; $[Yb^{3+}]/([Cr^{3+}] + [Yb^{3+}])$, with an estimated uncertainty of ±0.1%

**Variable-Temperature Photoluminescence (VTPL).** For VTPL measurements, single crystals were placed between two quartz disks and loaded into a closed-cycle helium cryostat, which was evacuated to pressures below $10^{-6}$ Pa. Samples were cooled to 4 K before measurements began. All variable-temperature series started at low temperature and progressed to higher temperatures. Samples were excited with a CW 660 nm (1.88 eV) diode at 4 mW/cm$^2$. Sample emission was collected, focused into a monochromator with a spectral bandwidth of 0.63 nm, and detected by a Hamamatsu InGaAs/InP liquid-nitrogen-cooled NIR photomultiplier. All spectra were corrected for instrument response. All luminescence spectra are displayed as photon counts versus energy (eV) following appropriate conversion.[57, 58]

**Variable-Temperature Time-Resolved Photoluminescence (VT-TRPL).** Time-resolved measurements were performed under the same conditions as used for CW PL measurements. An EKSPLA PL2230 Nd:YAG/YVO$_4$ laser with a pulse frequency of 50 Hz and a pulse duration of 20 ps was used to excite the samples at 532 nm (2.33 eV). Samples were excited with ~2 nJ/cm$^2$ per pulse. For decay traces, two emission energies were monitored: 1.2 eV and 1.15 eV for $Cr^{3+}$ and $Yb^{3+}$ + $Cr^{3+}$ emission, respectively, each with a spectral bandwidth of 8 meV. $Cr^{3+}$ PL decay signals were collected with a bin size of 5 ns, and $Yb^{3+}$ signals were collected with a bin size of 20 ns. The decay curves collected at 1.20 eV were fit using a bi-exponential function. The reported $\tau_{Cr}$ values correspond to the weighted average of biexponential decay times. To determine $\tau_{Yb}$, the data collected at 1.15 eV were considered. All $Cr^{3+}$ emission has fully decayed by ~15 μs, and the data from 15-200 μs thus corresponds solely to $Yb^{3+}$ emission. These latter data were fit to a biexponential and the weighted average of the two exponential decay times is reported as $\tau_{Yb}$. Note that at ~4 K, where $k_{Yb}^{rad}$ is determined, $Yb^{3+}$ PL decay is very nearly single-exponential, and a longer second component grows in at temperatures above ~50 K. Gated PL spectra were collected using a gated photon counter (Stanford Research Systems SR400) only integrating over 50 ns bin widths delayed by various times following the excitation pulse.

**Monte Carlo Simulations.** Code for Monte Carlo random-walk-type simulations was developed in python 3.10.12; see Figure S8 for flowchart. Simulations were performed as follows: For each $Yb^{3+}$ concentration, 250 independent trial crystal lattices were generated each having random distributions of the prescribed concentration of $Yb^{3+}$ ions. The size of the simulated crystal (2600 unit cells) was large enough that no simulated excitons hit the edge of any crystal. Lattice constants were taken from ICSD Coll. Code 251655.[59] On each trial lattice, 250 excitation trials were performed. Each excitation trial began with an exciton on a central $Cr^{3+}$ ion. For this initial state, four potential paths are possible: hopping to a neighboring $Cr^{3+}$ ion, emitting from the current ion, hoping to a neighboring $Yb^{3+}$ ion if one exists, or non-radiative relaxation. Each of these transitions from the initial state $i$ to another state $j$ was given a rate constant, $k_{ij}$, that characterizes its probability per unit time, assuming first-order exponential statistics. The rate constants for Cr-Cr hopping and non-radiative relaxation are modified by temperature (see main text). Which stochastic event occurs is determined by randomly generating a number $r$ ($0 < r \leq 1$) for each potential event, from which a dwell time of $t_{dwell} = -(1/k_{ij})\ln(r)$ describes how long state $i$ remains before transitioning to $j$. The event with the minimum $t_{dwell}$ is identified as the determined event, and the system moves to state $j$ with the recorded time. The process then repeats from this new initial state (for more details see Voter[60]), and continues to repeat after each transition until the trial ends due to emission, non-radiative relaxation, or annihilation of the exciton by $Yb^{3+}$. The entire sequence was repeated for a total of 62,500 excitons with the results aggregated over all random lattices. The emission events were binned in 50 ns intervals. Simulated decay curves were fit to determine the simulated decay times. The spatial locations of all $N$ exciton emission events were recorded, with each individual exciton diffusion length $x_i$ corresponding to the distance between the starting and ending position of that exciton. The aggregate diffusion length $L_D$ was calculated as $L_D = \sqrt{1/N \sum_i x_i^2}$ .

## Supporting Information

## Acknowledgments

This work was primarily supported by the U.S. National Science Foundation, Division of Materials Research (solid state and materials chemistry), through award DMR-2404703. Early stages of this work were supported by the U.S. National Science Foundation *via* Materials Research Science and Engineering Center (MRSEC) award DMR-2308979.

## Author Information

### Corresponding Author
**Daniel R. Gamelin** - *Department of Chemistry, University of Washington, Seattle, Washington 98195-1700, United States*; orcid.org/0000-0003-2888-9916;
Email: gamelin@chem.washington.edu

**Authors**

**Kimo Pressler** - *Department of Chemistry, University of Washington, Seattle, Washington 98195-1700, United States;* orcid.org/0000-0003-2788-1592

**Table of Contents Graphic**

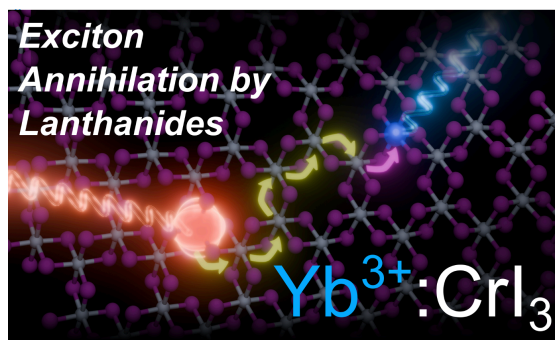

# Exciton Annihilation by Lanthanide Dopants: An Atomic Probe of Sub-Diffraction Exciton Diffusion in Ferromagnetic CrI$_3$

*Kimo Pressler, Daniel R. Gamelin\**
*Department of Chemistry, University of Washington, Seattle, WA 98195, United States*
Email: *gamelin@uw.edu*

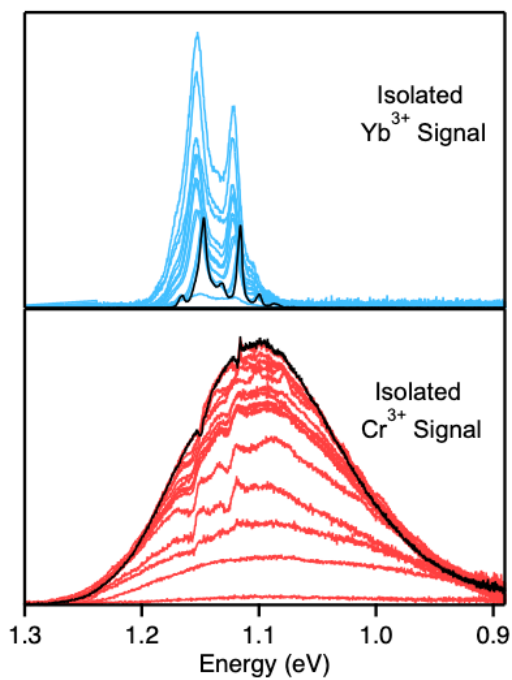

**Figure S1.** VTPL spectra of the same 0.2% Yb$^{3+}$:CrI$_3$ sample described in the main text, deconvoluted into Yb$^{3+}$ (blue) and Cr$^{3+}$ (red) components. To separate the two PL components, the 18.6% Yb$^{3+}$:CrI$_3$ PL spectrum, which features only Yb$^{3+}$ PL, was scaled and subtracted from the 0.2% spectra, giving the two data sets plotted here. The integrated intensities of these spectra are plotted in Figure 4e,f of the main text. The same deconvolution procedure was performed for the 1.6% Yb$^{3+}$:CrI$_3$ sample.

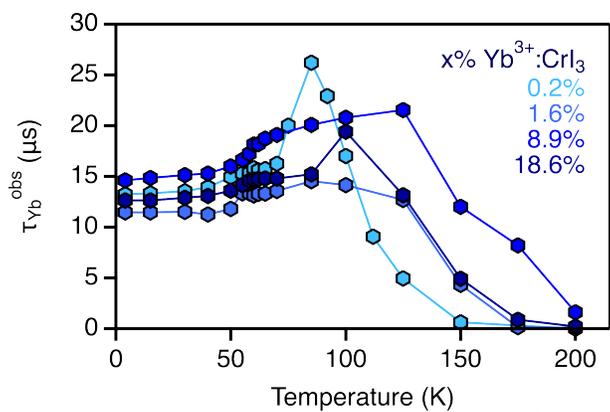

**Figure S2**. Plot of $Yb^{3+}$ PL decay times *vs* temperature for $x$% $Yb^{3+}$:$CrI_3$ ($x$ = 0.2, 1.6, 8.9, 18.6), collected with $CrI_3$ photoexcitation.

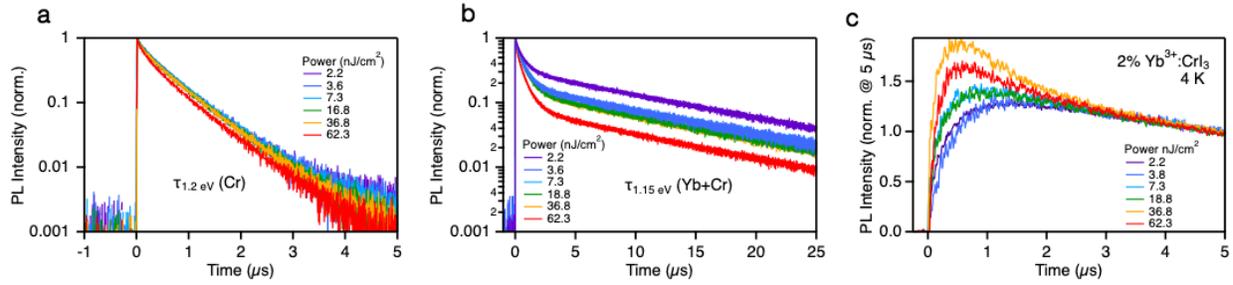

**Figure S3.** Normalized PL decay curves for 1.6% $Yb^{3+}$:$CrI_3$ measured as a function of excitation power. PL was measured at **(a)** 1.20 eV, representing $CrI_3$ emission ($Cr^{3+}$), and **(b)** 1.15 eV, representing a superposition of $CrI_3$ and $Yb^{3+}$ emission ($Yb^{3+}$ + $Cr^{3+}$), respectively. With increasing excitation power, the relative contribution of the $Yb^{3+}$ emission decreases, likely due to $Yb^{3+}$ PL saturation. **(c)** $Yb^{3+}$-only PL decay curves for each excitation power density, found by scaling and subtracting the data in panel (a) from the data in panel (b) at the same power (see main text). The rise time for the $Yb^{3+}$ PL decreases with increasing excitation power, consistent with increased probabilities of $Cr^{3+}$ exciton capture by $Yb^{3+}$ at increased excitation densities. All time-resolved PL data reported in the main text were collected using the lowest power density of 2 $nJ/cm^2$, corresponding to one absorbed photon per ~7 × $10^6$ $Cr^{3+}$ ions (see Excitation Density Calculations below).

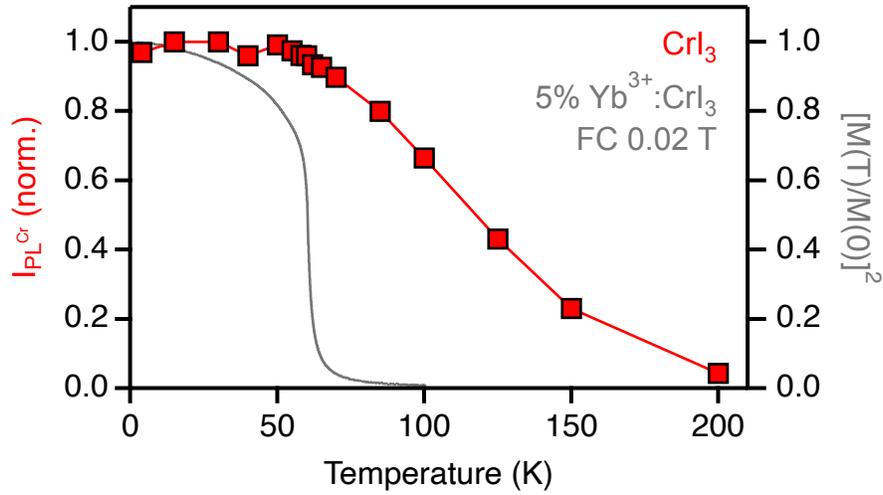

**Figure S4.** Comparison of the integrated PL intensity of undoped $CrI_3$ (red) with the lattice magnetization measured in using a magnetic field of 0.02 T applied parallel to the crystal's c-axis. Magnetization is plotted as the infinite spin-correlation function, represented by $[(M(T)/M(0)]^2$. The lattice magnetization decreases rapidly before reaching near zero at $T_C = 61$ K, whereas $I_{PL}$ shows a different and more graduate change extending to much higher temperatures. This comparison provides no indication of correlation between PL intensity and spin order. The magnetization data are taken from Pressler *et al.*[1]

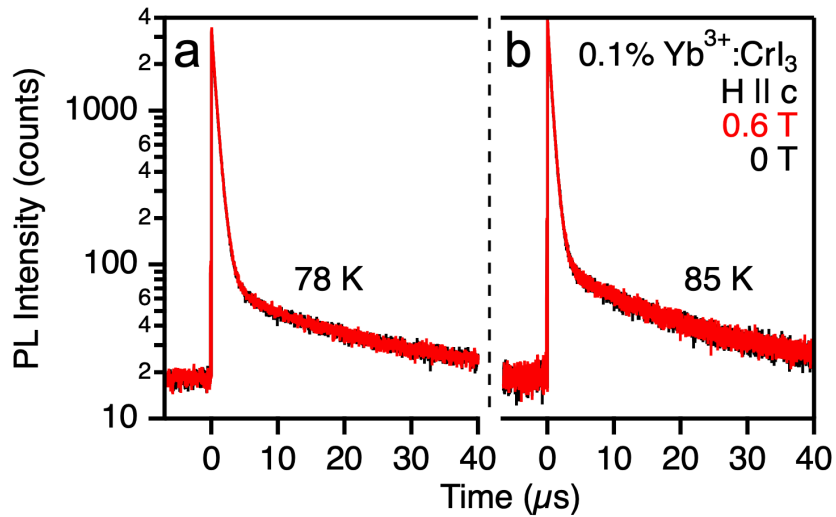

**Figure S5. (a)** 78 K and **(b)** 85 K PL decay curves of 1% $Yb^{3+}$:$CrI_3$ measured at 1.20 eV showing a superposition of $Cr^{3+}$ and $Yb^{3+}$ PL signals. Data were collected using a 40 ps, 50 Hz pulsed laser excitation source at 2.33 eV. The black traces correspond to no applied external field, while the red traces correspond to a 0.6 T field applied parallel to the crystal's c-axis. At 78 K (just above $T_C$), the change in lattice magnetization is appreciable with 0.6 T,[2] but there is no discernible difference in the decay dynamics upon application of the external field. These data show no indication of any sensitivity of exciton diffusion to lattice spin ordering. A similar result is obtained at 85 K.

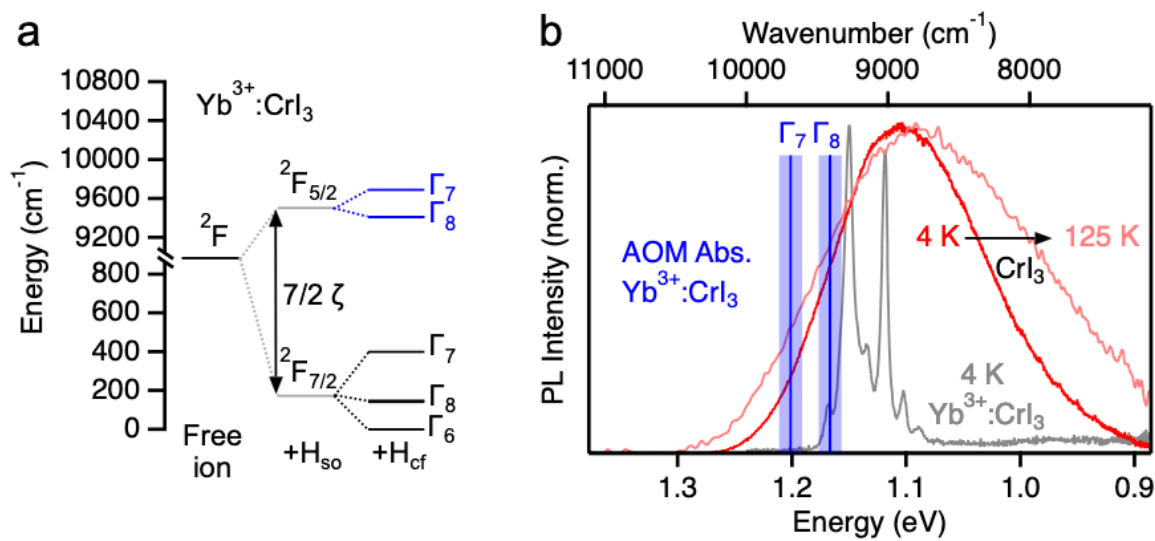

**Figure S6. (a)** *f*-shell spin-orbit and crystal-field splittings for $Yb^{3+}$ dopants in $CrI_3$, calculated using the Angular Overlap Model (AOM). See Snoeren[3] for calculation details. The calculated energies of the $\Gamma_7$ and $\Gamma_8$ excited crystal-field states are shown in blue in panel b. **(b)** PL spectrum of undoped $CrI_3$ at 4 K (red) and 125 K (pink) and of 18.6% $Yb^{3+}$:$CrI_3$ at 4 K (grey), plotted together with the calculated energies of the $Yb^{3+}$ $\Gamma_7$ and $\Gamma_8$ excited states (blue). The $Yb^{3+}$ excited states lie in a region of broad donor $Cr^{3+}$ PL at both low and high temperatures, resulting in resonant $Cr^{3+} \rightarrow Yb^{3+}$ energy transfer with a spectral overlap factor that is largely temperature-independent.

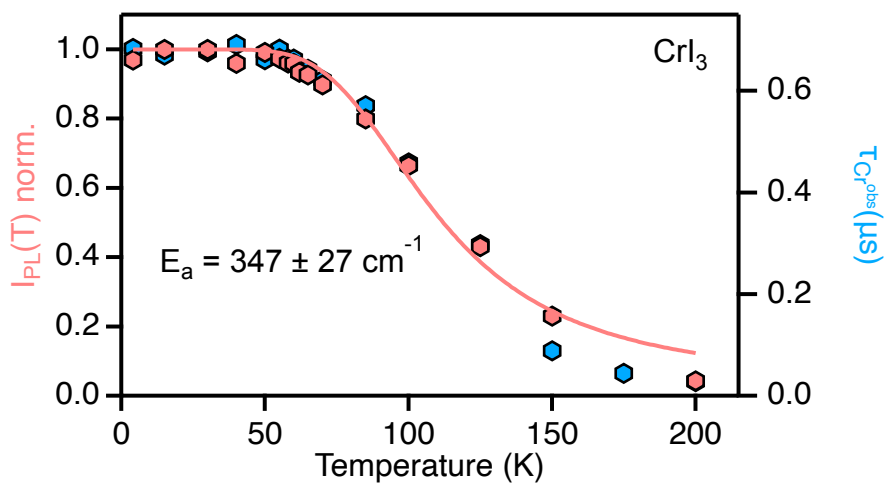

**Figure S7.** CrI$_3$ integrated PL intensities (red) and Cr$^{3+}$ PL decay times (blue) plotted *vs* temperature. The solid red curve plots the best fit of the integrated PL data to an Arrhenius function: $I_{PL}(T) = (I_0/(1+a*\exp(-E_a/(k_BT))))$, yielding $E_a$ = 347 ± 27 cm$^{-1}$. The two data sets show the same temperature dependence.

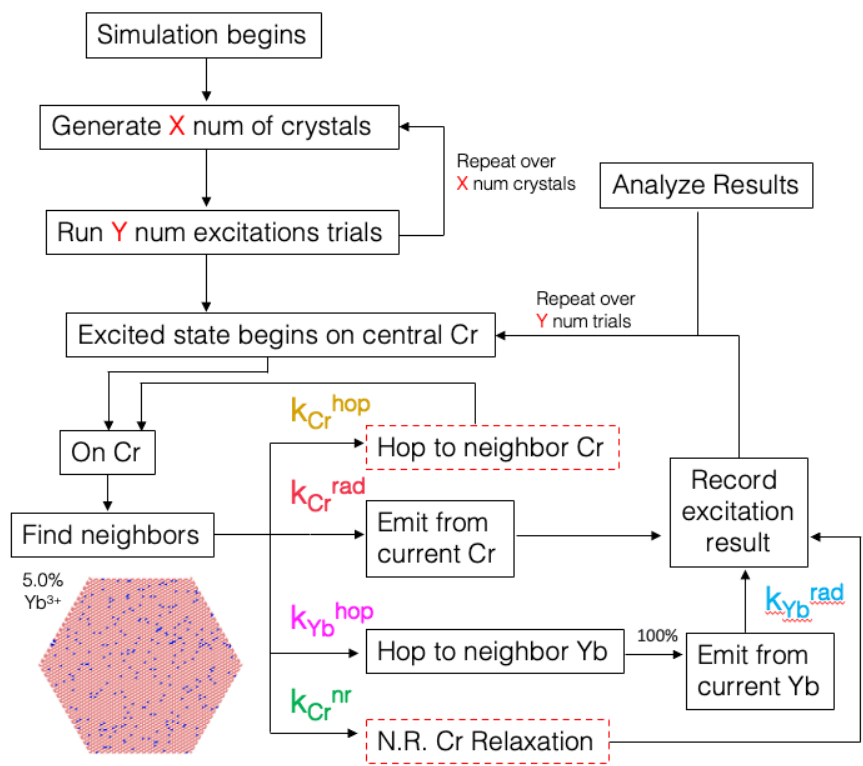

**Figure S8.** Flowchart of the Monte Carlo simulation, with a single simulated lattice depicted for reference. See Methods for details. The processes outlined in red are temperature dependent. The data shown in the main text are from simulations of over 250 crystals at every given doping level, each with 250 excitation trials at every given temperature, for a total of 62,500 simulated excitons for each doping level at each temperature.

**Table S1.** Rate constants and their values determined by Kinetic Monte Carlo simulations.

| Simulation parameter | 4 K rate constant (s$^{-1}$) | 200 K rate constant (s$^{-1}$) |
|---|---|---|
| $k_{Cr}^{rad}$ | 1.9 x 10$^6$ | temperature independent |
| $k_{Cr}^{nr}$ | ~0 | 9.9 x 10$^6$ |
| $k_{Cr}^{hop}$ | 1.0 x 10$^7$ | 6.8 x 10$^7$ |
| $k_{Yb}^{rad}$ | 1.1 x 10$^7$ | temperature independent |
| $k_{Yb}^{hop}$ | 2.0 x 10$^8$ | temperature independent |

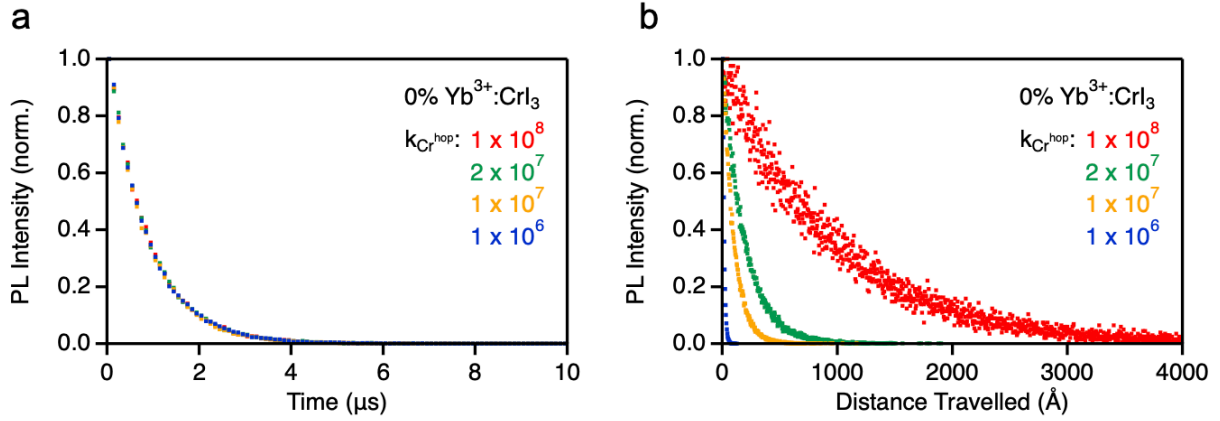

**Figure S9. (a)** Simulated PL decay curves for undoped $CrI_3$ at 4 K, calculated for various values of $k_{Cr}^{hop}$. The PL decay curves are all indistinguishable because emission events from different $Cr^{3+}$ ions are statistically independent of one another, and the ensemble decay therefore does not depend on exciton hopping. **(b)** The same data as in panel (a), but now plotted *vs* the exciton's distance traveled. With increasing $k_{Cr}^{hop}$, the exciton travels further from its point of inception. This dependence allows determination of $k_{Cr}^{hop}$ *via* simulation of PL data for different $Yb^{3+}$ doping levels. See main text and Figure 6b.

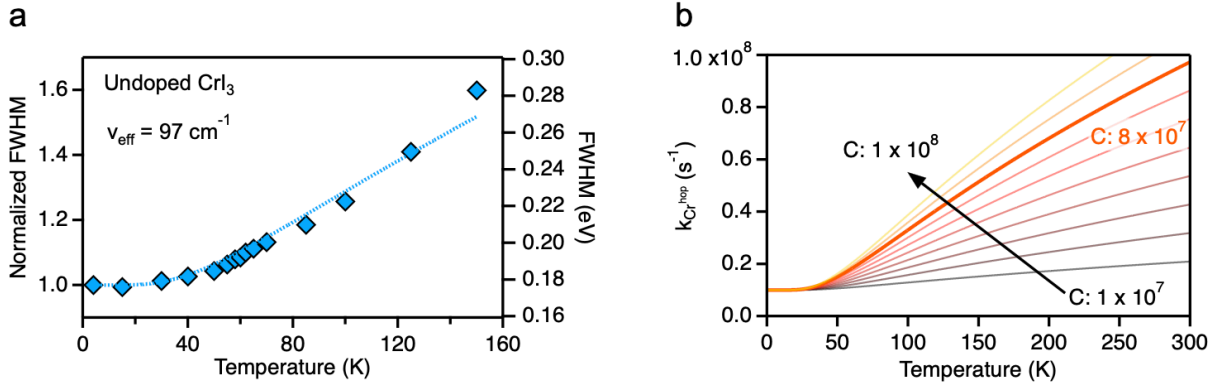

**Figure S10. (a)** Temperature dependence of the FWHM of the undoped CrI$_3$ PL, normalized to its low-temperature value. Following harmonic oscillator theory and assuming linear coupling to a single distortion coordinate, the PL full width at half maximum (FWHM) can be fit to eq S1.[4] The best-fit value for $v_{eff}$ (dotted curve) was found to be 97 cm$^{-1}$. **(b)** Plot of eq S2 (eq 11 of the main text) for various values of the scaling parameter *C*. The bolded orange line ($C = 8 \times 10^7$ s$^{-1}$) corresponds to the value that best reproduces the experimental data (Figure 7 of the main text).

$$FWHM(T) = FWHM(0)\left[\coth\left(\frac{v_{eff}}{2k_BT}\right)\right]^{1/2} \tag{S1}$$

$$k_{Cr}^{hop}(T) = C\left[\left[\coth\left(\frac{v_{eff}}{2k_BT}\right)\right]^{\frac{1}{2}} - 1\right] + k_{Cr}^{hop}(0) \tag{S2}$$

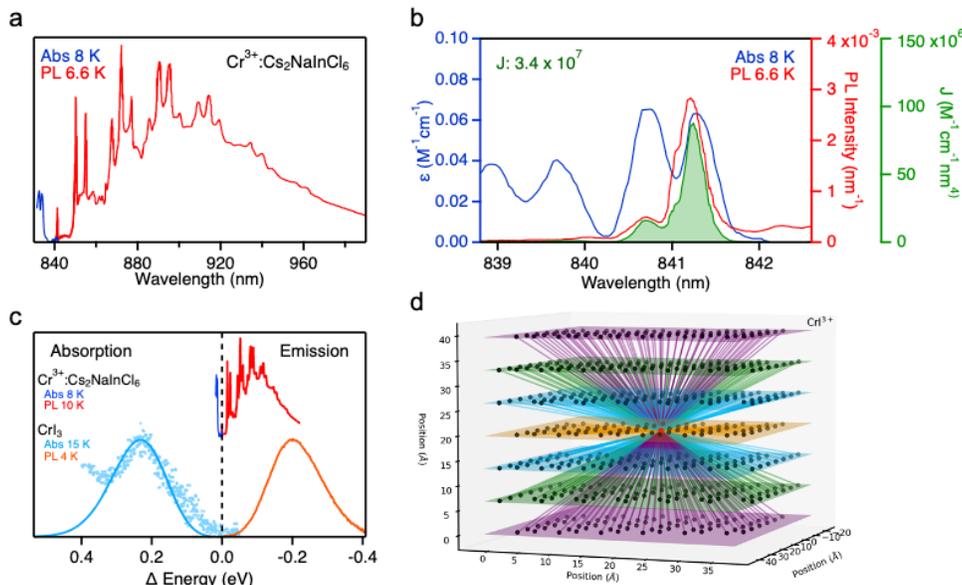

**Figure S11. (a)** Low-temperature PL (red) and absorption (blue) spectra of $Cr^{3+}$:$Cs_2NaInCl_6$, digitized from Güdel and Snellgrove.[5] **(b)** The overlap region of the absorption (blue) and PL (red) spectra of $Cr^{3+}$:$Cs_2NaInCl_6$, and the calculated spectral overlap (green). **(c)** Comparison of the low-temperature spectral overlap in $Cr^{3+}$:$Cs_2NaInCl_6$ and $CrI_3$. For $CrI_3$, the absorption is digitized from Seyler *et al.* and the PL is from this work.[6] **(d)** Depiction of the positions of 557 $Cr^{3+}$ ions within a 7-layer $CrI_3$ crystal of 32 unit cells. Atomic positions were determined from the low-temperature rhombohedral R-3H crystal structure ($a = b = 6.867$ Å, $c = 19.807$ Å, ICSD Coll. Code 251655).[7] The colored lines represent the distances from a central $Cr^{3+}$ ion (red) to all other $Cr^{3+}$ ions. These distances were used calculate the FRET rate constants shown in Table S2 as described in the following section.

**Table S2.** Calculated ET rate constants.

| Energy transfer process | Rate constant (s$^{-1}$) |
|---|---|
| Intra-layer Dexter NN (3.69 Å) | 1.00 x 10$^7$ |
| Intra-layer FRET NN (3.69 Å) | 7.73 x 10$^4$ |
| Intra-layer FRET | 1.41 x 10$^5$ |
| 1$^{st}$ Inter-layer FRET NN (6.87 Å) | 1.87 x 10$^3$ |
| 1$^{st}$ Inter-layer FRET | 1.32 x 10$^4$ |
| 2$^{nd}$ Inter-layer FRET | 7.73 x 10$^2$ |
| 3$^{rd}$ Inter-layer FRET | 1.25 x 10$^2$ |
| Total 1$^{st}$ - 3$^{rd}$ Inter-layer FRET | 1.41 x 10$^4$ |
| Total Intra- + Inter-layer FRET | 1.56 x 10$^5$ |

**Calculation of $Cr^{3+}$-$Cr^{3+}$ Energy Transfer and FRET Rate Constants**

For $k_{Cr}^{hop}$, two possible mechanisms of energy transfer are considered: through-space multipolar Förster resonant energy transfer (FRET) and through-bond Dexter energy transfer (DET), both of which may be present. The rate constants for each process are given by eq S3 and S4, respectively.

$$k_{FRET} = 0.02108 \frac{\theta^2 Q_D J}{n^4} \left(\frac{1}{r}\right)^6 \frac{1}{\tau_D} \tag{S3}$$

$$k_{DET} = KJ' exp\left[\frac{-2r}{L}\right] \tag{S4}$$

For FRET, $\theta^2$ is the donor acceptor dipole orientation factor, $Q_D$ is the quantum yield of the donor PL, $J$ is the spectral overlap integral defined in eq S5, $n$ is the refractive index of the lattice (1.94 at 1000 nm in $CrI_3$),[8] $\tau_D$ is the donor PL decay time (837 ns in $CrI_3$), and $r$ is the distance between the donor and acceptor. For DET $K$ is a constant describing the magnitude of the donor-acceptor electronic coupling, $J'$ is the spectral overlap integral defined in eq S6, $r$ is the separation between donor and acceptor, and $L$ is the sum of their van der Waals radii.

$$J = \int \bar{f}_D(\lambda)\varepsilon_A(\lambda)\lambda^4 d\lambda \tag{S5}$$

$$J' = \int \bar{f}_D(\lambda)\bar{\varepsilon}_A(\lambda)d\lambda \tag{S6}$$

Here $\bar{f}_D(\lambda)$ represents the area-normalized donor emission spectrum, $\varepsilon_A(\lambda)$ the acceptor molar extinction coefficient, and $\bar{\varepsilon}_A(\lambda)$ the area-normalized acceptor molar extinction coefficient.

Although both DET and FRET contain a spectral overlap component, an important distinction lies in how the spectral overlap integrals *(J* or *J')* are calculated. For FRET, *J* is calculated as the overlap integral of the donor luminescent spectrum normalized to a unit area, and the acceptor extinction, whereas in *J'* both the donor luminescence spectrum and acceptor extinction spectrum are normalized by area.[9] Accordingly, the rate of dipole-dipole FRET depends on the oscillator strengths of both the donor and acceptor transitions, whereas the rate of DET does not depend on these oscillator strengths and only requires the existence of such states resonant in energy.

Because all variables in eq S3 can be determined experimentally, we sought to estimate the contribution of FRET to $Cr^{3+}$-$Cr^{3+}$ ET at low temperature. In $CrI_3$, strong electron-phonon coupling and a large ~400 meV Stokes shift result in extremely low *J*, and all overlap results solely from the weak electronic origins (see Figure 4g). To estimate *J*(4K) and FRET(4K) in $CrI_3$, we used the reference compound $Cr^{3+}$:$Cs_2NaInCl_6$,[5] which displays $Cr^{3+}$ emission from pseudo-octahedral $CrCl_6^3$ and has resolvable electronic origins in both its absorption and PL spectra. Figure S11c compares the relevant spectra for $CrI_3$ and $Cr^{3+}$:$Cs_2NaInCl_6$. Using eq S3 along with the high-resolution PL (red trace in S11b) and extinction (blue trace in S11b) spectra, *J* was estimated to be 3.4 x $10^7$ $M^{-1}cm^{-1}nm^4$ in $Cr^{3+}$:$Cs_2NaInCl_6$. Because of the greater Stokes shift, this value is considered to provide an upper bound for the value of *J* in $CrI_3$.

To assess the importance of FRET in $CrI_3$, distances were determined from a central $Cr^{3+}$ ion to 557 $Cr^{3+}$ ions within a 7-layer $CrI_3$ crystal consisting of 32 unit cells (Figure S11d). These distances were then used to calculate individual FRET rate constants from that central ion to each other $Cr^{3+}$ ion of this crystal, and the results are summarized in Table S2. In addition to using an upper bound for *J* in $CrI_3$ (see above), these FRET calculations used favorable values of 1.0 for $Q_D$ and 2/3 for $\theta^2$. The results in Table S2 thus represent upper bounds for the FRET rate constants.

As seen in Table S2, all calculated FRET rate constants are orders of magnitude smaller than the nearest neighbor hopping rate constant of $k_{Cr}^{hop}$ = 1.00 x 10$^7$ s$^{-1}$ determined in the main text. This comparison supports the conclusion that energy migration in CrI$_3$ proceeds through intralayer DET.

**Excitation Densities**

To estimate the experimental excitation densities in CrI$_3$, we first find the number of photons hitting the sample in each excitation pulse. The sample was excited by a 532 nm laser with energy density of 2.0 nJ/cm$^2$ over a spot size with a diameter of 1.0 mm. The energy of one 532 nm excitation photon is given by eq S7:

$$E = \frac{hc}{\lambda} = \frac{(6.63 \times 10^{-34}\ J \cdot s)(2.998 \times 10^8\ m/s)}{532 \times 10^{-9}\ m} = 3.74 \times 10^{-19}\ J/photon \quad (S7)$$

The density of photons per excitation pulse is thus:

$$\frac{2.0 \times 10^{-9}\ J/cm^2}{3.74 \times 10^{-19}\ J/photon} = 5.3 \times 10^9\ photons/cm^2 \quad (S8)$$

Over the excitation spot size, this gives:

$$5.3 \times 10^9 \frac{photons}{cm^2} * \pi(0.050\ cm)^2 = 4.2 \times 10^7\ photons \quad (S9)$$

The penetration depth (z) of the excitation pulse corresponds to the distance at which the light intensity falls to 1/e (~37%) of its initial value and can be found using the material's extinction (attenuation) coefficient $\kappa(\lambda)$ at the light's wavelength. Using the value of 1.19 cm$^{-1}$ for $\kappa$ at 532 nm in CrI$_3$,[10] z is given by eq 10:

$$I(z) = I_0\ e^{-4\pi\kappa(\lambda)z/\lambda_0} \rightarrow \frac{I(z)}{I_0} = \frac{37}{100} = e^{-4\pi\kappa(\lambda)z/\lambda_0} \rightarrow z = \frac{\ln(\frac{37}{100})}{-4\pi\kappa(\lambda)/\lambda_0} = 3.5 \times 10^{-6} cm \quad (S10)$$

This penetration depth corresponds to ~50 layers of CrI$_3$. Given the excitation spot size, this gives a volume of:

$$V = \pi r^2 l = \pi(0.050\ cm)^2(3.5 \times 10^{-6} cm) = 2.7 \times 10^{-8} cm^3 \quad (S11)$$

From the density of CrI$_3$, this volume corresponds to $1.9 \times 10^{14}$ total Cr$^{3+}$ ions:

$$4.94 \frac{g}{cm^3} * \frac{6.02 \times 10^{23}\ ions}{432.96\ g} * 2.7 \times 10^{-8} cm^3 = 1.9 \times 10^{14}\ ions \quad (S12)$$

63% of the photons are absorbed within this volume, giving:

$$\frac{1.9 \times 10^{14}\ Cr\ ions}{(0.63) * 4.2 \times 10^7\ photons} = 7.2 \times 10^6\ Cr\ ions/exciton \quad (S13)$$

At such low excitation densities, exciton-exciton annihilation is deemed negligible.